% =======================LETRAS HUECAS============================
\newfam\msbfam
\font\twlmsb=msbm10 at 12pt
\font\eightmsb=msbm10 at 8pt
\font\sixmsb=msbm10 at 6pt
\textfont\msbfam=\twlmsb
\scriptfont\msbfam=\eightmsb
\scriptscriptfont\msbfam=\sixmsb
\def\cj{\fam\msbfam}
\def\A{{\cj A}}

\def\R{{\cj R}}

\centerline{\bf Poincar\'e gauge invariance of general relativity and Einstein-Cartan theory} 

\

\

\centerline{G. D'Olivo and M. Socolovsky}

\

\centerline{\it  Instituto de Ciencias Nucleares, Universidad Nacional Aut\'onoma de M\'exico}
\centerline{\it Circuito Exterior, Ciudad Universitaria, 04510, M\'exico D. F., M\'exico} 

\

{\bf Abstract} {\it We present a simple proof of the Poincar\'e gauge invariance of general relativity and Einstein-Cartan theory, in the context of the corresponding bundle of affine frames.}

\


\

{\bf 1. Introduction}

\

It is well understood that general relativity (GR) and its extension with torsion, the Einstein-Cartan theory (E-C), are invariant under internal local Lorentz transformations (${\cal L}_4$), the spin connection $\omega_{\mu ab}$ and the tetrads ${e_\mu}^a$ (coframes) (or rather the displaced fields ${B_\mu}^a={\delta_\mu}^a-{e_\mu}^a$) being respectively the rotational and translational gravitational gauge potentials (Hehl, 1985; Hayashi, 1977). Then, the group of symmetry of both theories is the semidirect sum ${\cal L}_4\odot {\cal D}$, with ${\cal D}$ the group of general coordinate transformations (O'Raifeartaigh, 1997). 

\

However, the internal symmetry group is in fact larger, since translations ${\cal T}_4$ are {\it naturally} included, leading to ${\cal P}_4={\cal T}_4\odot{\cal L}_4$, the Poincar\'e group. Then, the total symmetry of GR and E-C, as gauge theories, turns out to be ${\cal P}_4\odot{\cal D}$ (Feynman, 1963; Hehl et al, 1976; Mc Innes, 1984; Hammond, 2002). The problem with the proof of this fact has been, historically, the apparent difficulty with the treatment of translations as part of the gauge group, that is, as {\it vertical} transformations of a bundle. If for a translation one writes $x^\mu \to x^{\prime\mu}=x^\mu +\xi(x)$, one is {\it not} considering it as a ${\cal P}_4$-gauge transformation, but instead as an element of ${\cal D}$. The appropiate treatment of gauge translations is in the framework of the {\it bundle of Poincar\'e frames} over space-time, ${\cal F}^P_{M^4}$: ${\cal P}_4\to A^PM^4\buildrel{\pi_P}\over\longrightarrow M^4$. 

\

This has been discussed by several authors (Smrz, 1977; Gronwald, 1997, 1998), and it is the purpose of this note to present an even simpler proof of this fact. On the one hand, at the global level, we show, using general theorems of connections (Kobayashi-Nomizu, 1963), that there is a 1-1 correspondence between affine Poincar\'e connections $\omega_P$ in ${\cal F}^P_{M^4}$ and pairs $(\theta_L,\omega_L)$, with $\theta_L$ the canonical form and $\omega_L$ a connection on the bundle of Lorentz frames ${\cal F}^L_{M^4}$: ${\cal L}_4\to F^LM^4\buildrel{\pi_L}\over\longrightarrow M^4$. On the other hand, locally, we show the invariance under ${\cal P}_4$-gauge transformations of the Einstein-Hilbert action for pure gravity, and the Dirac-Einstein action for the coupling of gravity to the Dirac field.

\

In section 2, we describe basic features of a $U_4$ space-time. In section 3, in the language of tetrads and spin connection, we review the E-C equations for pure gravity and for gravity coupled to the Dirac field. Lorentz and Poincar\'e invariance are discussed and proved in sections 4 and 5, respectively. Finally, in section 6, we discuss the nature of a shifted tetrad field, and comment on the difficulty of interpreting the theory in terms of an interaction tetrads-spin connection.

\

{\bf 2. The space-time}

\

We assume that space-time is a 4-dimensional Lorentzian manifold $M^4$ with a connection $\Gamma$ compatible with the metric i.e. $D_\mu^\Gamma g_{\nu\rho}=0$, but not necessarily symmetric: a $U_4$ space-time. Then, $\Gamma^\alpha_{\nu\mu}={(\Gamma_{LC})}^\alpha_{\nu\mu}+K^\alpha_{\nu\mu}$, where $\Gamma_{LC}$ is the Levi-Civita connection with coordinate components $(\Gamma_{LC})^\alpha_{\nu\mu}={{1}\over{2}}g^{\alpha\sigma}(\partial_\nu g_{\mu\sigma}+\partial_\mu g_{\nu\sigma}-\partial_\sigma g_{\nu\mu})$, and $K^\alpha_{\nu\mu}={(K_A)}^\alpha_{\nu\mu}+{(K_S)}^\alpha_{\nu\mu}$ is the contortion tensor, where ${(K_A)}^\alpha_{\nu\mu}=T^\alpha_{\nu\mu}=-T^\alpha_{\mu\nu}={{1}\over{2}}(\Gamma^\alpha_{\nu\mu}-\Gamma^\alpha_{\mu\nu})=\Gamma^\alpha_{[\mu,\nu]}$ is the {\it torsion} tensor, and $K_S$, its symmetric part, has components ${(K_S)}^\alpha_{\mu\nu}=g^{\alpha\rho}(T^\lambda_{\rho\mu}g_{\lambda\nu}+T^\lambda_{\rho\nu}g_{\lambda\mu}).$ 

\

In terms of the tetrads $e_a={e_a}^\mu\partial_\mu$ and their dual coframes $e^a={e_\mu}^adx^\mu$, obeying ${e_a}^\mu {e_\mu}^b=\delta^b_a$ and ${e_a}^\mu {e_\nu}^a=\delta^\mu_\nu$, and the spin connection 1-form ${\omega^a}_b=\omega_{\mu b}^a dx^\mu$, $\Gamma$ is given by $\Gamma_{\mu\lambda}^\sigma={e_a}^\sigma\partial_\mu{e_\lambda}^a+{e_c}^\sigma{e_\lambda}^a\omega_{\mu a}^c$ with inverse $\omega^c_{\mu a}={e_\rho}^c\partial_\mu{e_a}^\rho +{e_\rho}^c{e_a}^\nu\Gamma^\rho_{\mu\nu}$ (Carroll, 2004). For the metric, one has $g_{\mu\nu}(x)=\eta_{ab}{e_\mu}^a(x){e_\nu}^b(x)$, where $x\in M^4$ and $\eta_{ab}=diag(1,-1,-1,-1)$ is the Lorentz metric. Each metric $g_{\mu\nu}$ is in 1-1 correspondence with an equivalence class of frames $[{e_a}^\mu]$: if ${e_c}^{\prime\mu}$ is in the class, then ${e_a}^\mu={h_a}^c{e_c}^{\prime\mu}$ with ${h_a}^c$ in the Lorentz group ${\cal L}_4$; for the coframes ${e_\mu}^a={e_\mu}^{\prime c}{h_c^{-1}}^a$. Thus, the ${e_a}^\mu$'s and the ${e_\mu}^a$'s are both Lorentz vectors in the internal or gauge (latin) indices, and respectively vectors and 1-forms in the local coordinate (world) indices. The metric character of the connection implies $\omega_{ab}=-\omega_{ba}$ (for latin indices, $X_a=\eta_{ab}X^b$ and $X^b=\eta^{ba}X_a$). The torsion and the curvature of the connection are given by $T^a=de^a+{\omega^a}_b\wedge e^b={{1}\over{2}}{T^a}_{\mu\nu}dx^\mu\wedge dx^\nu$ with ${T^a}_{\mu\nu}=\partial_\mu {e_\nu}^a -\partial_\nu {e_\mu}^a +\omega^a_{\mu b}{e_\nu}^b-\omega^a_{\nu b}{e_\mu}^b,$ and ${R^a}_b=d{\omega^a}_b+{\omega^a}_c\wedge{\omega^c}_b={{1}\over{2}}{R^a}_{b\mu\nu} dx^\mu\wedge dx^\nu$ with $R^a_{b\mu\nu}=\partial_\mu\omega^a_{\nu b}-\partial_\nu\omega^a_{\mu b}+\omega^a_{\mu c}\omega^c_{\nu b}-\omega^a_{\nu c}\omega^c_{\mu b}.$ 

\

{\bf 3. Einstein-Cartan equations}

\

Consider first the case of {\it pure gravity} ("vacuum''). The Einstein-Hilbert action is $$S_G=\int d^4x \ eR \eqno{(1)}$$ where $e=\sqrt{-detg_{\mu\nu}}=det({e_\nu}^a)$, and for the Ricci scalar one has $$R=\eta^{bc}{R^a}_{b\mu\nu}{e_a}^\mu{e_c}^\nu. \eqno{(2)}$$ Variation of $S_G$ with respect to the spin connection $\omega_{\mu b}^a$ and the tetrads ${e_a}^\mu$ lead, respectively, to the Cartan equation for torsion and to the Einstein equation:

\

$\delta_\omega S_G=0 \Longrightarrow$ $$T^\nu_{ac}+{e_a}^\nu T_c-{e_c}^\nu T_a=0 \eqno{(3)}$$ or $$T^\nu_{\rho\sigma}+\delta^\nu_\rho T_\sigma-\delta^\nu_\sigma T_\rho=0. \eqno{(3a)}$$ \ \ \ \ \ \ $\delta_eS_G=0 \Longrightarrow$ $${G^a}_\mu=0 \eqno{(4)}$$ with  $${G^a}_\mu={R^a}_\mu-{{1}\over{2}}R{e_\mu}^a, \eqno{(5)}$$ where ${R^a}_\mu=\eta^{ab}R_{b\mu}=\eta^{ab}{R^c}_{b\nu\mu}{e_c}^\nu$. In vacuum $R=0$, then $${R^a}_\mu=0. \eqno{(5a)}$$ In this case, {\it torsion vanishes}, since taking the $\nu-\sigma$ trace in (3a), for the torsion vector one obtains $T_\rho=T^\nu_{\rho\nu}=0$ and therefore, by (3a) again, $$T^\mu_{\nu\rho}=0. \eqno{(6)}$$ Thus, for the pure gravity case, E-C theory reduces to GR.

\

The coupling of gravity to Dirac fermions is described by the action $$S_{D-E}=k\int d^4x \ e L_{D-E}=k\int d^4x \ e \ ({{i}\over{2}}(\bar{\psi}\gamma^a(D_a\psi)-(\bar{D}_a\bar{\psi})\gamma^a\psi ) -m\bar{\psi}\psi) \eqno{(7)}$$ where $$D_a\psi=(e_a-{{i}\over{4}}\omega_{abc}\sigma^{bc})\psi={e_a}^\mu(\partial_\mu-{{i}\over{4}}\omega_{\mu bc}\sigma^{bc})\psi={e_a}^\mu D_\mu\psi \eqno{(8)}$$ and $$\bar{D}_a\bar{\psi}=e_a\bar{\psi}+{{i}\over{4}}\omega_{abc}\bar{\psi}\sigma^{bc}={e_a}^\mu(\partial_\mu \bar{\psi}+{{i}\over{4}}\omega_{\mu bc}\bar{\psi}\sigma^{bc})={e_a}^\mu \bar{D}_\mu\bar{\psi} \eqno{(9)}$$ are the covariant derivatives of the Dirac field $\psi$ and its conjugate $\bar{\psi}=\psi^\dagger \gamma_0$ with respect to the spin connection, which give the {\it minimal coupling} between fermions and gravity. $\sigma ^{bc}={{i}\over{2}} [\gamma ^b,\gamma ^c]$, and the $\gamma ^a$'s are the usual numerical (constant) Dirac gamma matrices satisfying $\{\gamma ^a,\gamma ^b\}=2\eta ^{ab}I$, $\gamma ^{0\dag}=\gamma ^0$ and $\gamma ^{j\dag}=-\gamma ^j$. $k=-16\pi{{G}\over{c^4}}$. Variation with respect to the spin connection, $$\delta_\omega S_{D-E}={{k}\over{8}}\int d^4x \ e \ \bar{\psi}\{\gamma^\mu,\sigma^{bc}\}\psi\delta\omega_{\mu bc}={{k}\over{2}}\int d^4x \ e \ S^{\mu bc}\delta\omega_{\mu bc}$$ with $S^{\mu bc}={e_a}^\mu S^{abc}$, where $$S^{abc}={{1}\over{4}}\bar{\psi}\{\gamma^a,\sigma^{bc}\}\psi \eqno{(10)}$$ is the {\it spin density tensor} of the Dirac field. $S^{abc}$ is totally antisymmetric and therefore has 4 independent components: $S^{012}$, $S^{123}$, $S^{230}$ and $S^{301}$.  

\

Combining this result with the corresponding variation for the pure gravitational field, we obtain $$0=\delta_\omega (S_G+S_{D-E})=\int d^4x \ e \ \delta{\omega_\nu}^{ac}(T^\nu_{ac}+{e_a}^\nu T_c-{e_c}^\nu T_a+{{k}\over{2}}S^\nu_{ac}) \eqno{(11)}$$ and therefore $$T^\nu_{ac}+{e_a}^\nu T_c-{e_c}^\nu T_a=-{{k}\over{2}}S^\nu_{ac},$$ the {\it Cartan equation}. Multiplying by ${e_\rho}^a{e_\sigma}^c$ one obtains $$T^\nu_{\rho\sigma}+\delta^\nu_\rho T_\sigma-\delta^\nu_\sigma T_\rho=-{{k}\over{2}}S^\nu_{\rho\sigma} \eqno{(12)}$$ with $$S^\nu_{\rho\sigma}={{1}\over{4}}\bar{\psi}\{\gamma^\mu, \sigma_{\rho\sigma}\}\psi.$$ 

\

The solution of (12) gives the {\it torsion in terms of the spin tensor}: $$T^\nu_{\rho\sigma}={{8\pi G}\over{c^4}}(S^\nu_{\rho\sigma}+{{1}\over{2}}(\delta^\nu_\rho S_\sigma-\delta^\nu_\sigma S_\rho)) \eqno{(13)} $$ with $S_\rho=S^\nu_{\rho\nu}$. (In natural units, $G=c=\hbar=1$ and so $T^\nu_{\rho\sigma}=8\pi(S^\nu_{\rho\sigma}+{{1}\over{2}}(\delta^\nu_\rho S_\sigma-\delta^\nu_\sigma S_\rho))$.) 

\

Finally, variation with respect to the tetrads, $$\delta_e S_{D-E}=k\int d^4x \ e \ ({{i}\over{2}}(\bar{\psi}\gamma^a(D_\mu\psi)-(\bar{D}_\mu\bar{\psi})\gamma^a\psi)-{e_\mu}^aL_{D-E})\delta{e_a}^\mu.$$ For the Dirac fields which obey the equations of motion $${{\delta S_{D-E}}\over{\delta\bar{\psi}}}={{\delta S_{D-E}}\over{\delta\psi}}=0$$ i.e. $$i\gamma^a(\bar{D}_a\bar{\psi})+m\bar{\psi}=i\gamma^aD_a\psi-m\psi=0$$ the Dirac-Einstein lagrangian vanishes i.e. $L_{D-E}\vert_{eq. \ mot.}=0$. Then, combining this result with the corresponding variation for the pure gravitational field, $$0=\delta_e(S_G+S_{D-E})=\int d^4x \ e \ (2{R^a}_\mu-R{e_\mu}^a+k{{i}\over{2}}(\bar{\psi}\gamma^a(D_\mu\psi)-(\bar{D}_\mu\bar{\psi})\gamma^a\psi))\delta{e_a}^\mu, \eqno{(14)}$$ and from the arbitrariness of $\delta{e_a}^\mu$, $${R^a}_\mu-{{1}\over{2}}R{e_\mu}^a=-{{k}\over{2}}{T^a}_\mu \eqno{(15)}$$ with $${T^a}_\mu={{i}\over{2}}(\bar{\psi}\gamma^a(D_\mu\psi)-(\bar{D}_\mu\bar{\psi})\gamma^a\psi) \eqno{(16)}$$ the {\it energy-momentum tensor} of the Dirac field. Multiplying $(15)$ by ${e_a}^\nu$ one obtains $${R^\nu}_\mu-{{1}\over{2}}R\delta^\nu_\mu=-{{k}\over{2}}{T^\nu}_\mu \ \ or \ \ R_{\lambda\mu}-{{1}\over{2}}Rg_{\lambda\mu}=-{{k}\over{2}}T_{\lambda\mu}, \eqno{(15a)}$$ the {\it Einstein equation} in local coordinates.

\

{\it Note}: For $L_{D-E}$ one has $$L_{D-E}={e_a}^\mu{T^a}_\mu-m\bar{\psi}\psi$$ i.e. ${T^a}_\mu$ {\it couples to the tetrad}. On the other hand, $${T^a}_\mu={\theta^a}_\mu+\omega_{\mu bc}S^{abc}$$ where $${\theta^a}_\mu={{i}\over{2}}(\bar{\psi}\gamma^a\partial_\mu\psi-(\partial_\mu\bar{\psi})\gamma^a\psi)$$ is the {\it canonical energy-momentum tensor} of the Dirac field. Then, $$L_{D-E}={e_a}^\mu{\theta^a}_\mu+{e_a}^\mu\omega_{\mu bc}S^{abc}-m\bar{\psi}\psi={e_a}^\mu{\theta^a}_\mu+\omega_{abc}S^{abc}-m\bar{\psi}\psi.$$ So, ${\theta^a}_\mu$ couples to the tetrad while spin couples to the spin connection; moreover, since $S^{abc}$ is totally antisymmetric, {\it the Dirac field only interacts with the totally antisymmetric part of the connection}.

\

{\bf 4. Lorentz gauge invariance}

\

Under local Lorentz transformations ${h_a}^b(x)$, tetrads and coframes transform as indicated in section 2; as a consequence, the coordinate invariant volume element $d^4x \ e$ is also gauge invariant: in fact, $$g_{\mu\nu}(x)=\eta_{ab}{e_\mu}^a(x){e_\nu}^b(x)=\eta_{ab}{e^\prime_\mu}^c{h^{-1}_c}^a{e^\prime_\nu}^d{h^{-1}_d}^b={e^\prime_\mu}^c{e^\prime_\nu}^d{h^{-1}_c}^a\eta_{ab}{h^{-1}_d}^b={e^\prime_\mu}^c{e^\prime_\nu}^d\eta_{cd}=g^\prime_{\mu\nu}(x)$$ implies $e^\prime(x)=e(x)$, and since $x^{\prime\mu}=x^\mu$, then $d^4x \ e=d^4x^\prime \ e^\prime$. 

\

On the other hand, the transformation of the spin connection is given by $${\omega^c}_a={h_c}^d{\omega^{\prime r}}_d{h^{-1}_r}^c + (d{h_a}^d){h^{-1}_d}^c,\eqno{(16)}$$ which is not a Lorentz tensor. Its curvature, however, is a Lorentz tensor: $${R^a}_b={h_b}^d{h_c^{-1}}^a{R^{\prime c}}_d,\eqno{(17)}$$ and therefore the Ricci scalar is also gauge invariant: $$R={R^a}_b e_a\eta^{bc}e_c={h_b}^d{R^{\prime c}}_d{h^{-1}_c}^a{h_a}^f e^\prime_f \eta^{bg}{h_g}^l e^\prime_l={R^{\prime c}}_d\delta^f_c e^\prime_f\eta^{dl}e^\prime_l={R^{\prime c}}_d e^\prime_c\eta^{dl}e^\prime_l=R^\prime.\eqno{(18)}$$ Then, $S_G$ is Lorentz gauge invariant. (A direct and more explicit proof of the gauge invariance of $R$ is given in Appendix 1.)

\

The part of the action corresponding to the coupling of gravity to the Dirac field, $S_{D-E}$, is automatically local Lorentz invariant, since it is written in terms of the covariant derivatives $D_a\psi$ and $\bar{D}_a\bar{\psi}$. 

\

{\bf 5. Poincar\'e gauge invariance}

\

5.1. {\it Global analysis}

\

The affine group $GA_4(\R)=\{\pmatrix{g & \xi \cr 0 & 1\cr}, \ g\in GL_4(\R), \ \xi\in \R^4\}$ acts on the affine space $\A^4=\{\pmatrix{\lambda\cr 1 \cr}, \ \lambda\in \R^4\}$ in the form $$GA_4(\R)\times\A^4\to\A^4, \ (\pmatrix{g & \xi \cr 0 & 1\cr},\pmatrix{\lambda\cr 1 \cr})\mapsto \pmatrix{g\lambda+\xi \cr 1 \cr}.\eqno{(19)}$$ Then, one has the following diagram of short exact sequences (s.e.s.'s) of groups and group homomorphisms: $$\matrix{0 & \longrightarrow & \R^4 & \buildrel{\mu}\over\longrightarrow & GA_4(\R) & \matrix{\buildrel{\nu}\over\longrightarrow\cr\buildrel{\rho}\over\longleftarrow\cr} & GL_4(\R) & \longrightarrow & 0 \cr & & Id\uparrow & & \uparrow\iota & & \uparrow\iota & & & \cr 0 & \longrightarrow & \R^4 & \buildrel{\mu\vert}\over\longrightarrow & {\cal P}_4 & \matrix{\buildrel{\nu\vert}\over\longrightarrow\cr\buildrel{\rho\vert}\over\longleftarrow\cr} & {\cal L}_4 & \longrightarrow & 0 \cr}$$ with $\mu(\xi)=\pmatrix{I_4 & \xi \cr 0 & 1 \cr}$ and $\nu(\pmatrix{g & \lambda \cr 0 & 1})=g$. $\mu$ is 1-1, $\nu$ is onto, and $ker(\nu)=Im(\mu)=\{\pmatrix{I_4 & \xi \cr 0 & 1 \cr}, \ \xi\in \R^4\}$. We have also restricted $\mu$ and $\nu$ (respectively $\mu\vert$ and $\nu\vert$) to the connected components of the Poincar\'e (${\cal P}_4$) and Lorentz (${\cal L}_4$) groups. Both s.e.s.'s split, i.e. there exists the group homomorphism $\rho:GL_4(\R)\to GA_4(\R)$, $g\mapsto\rho(g)=\pmatrix{g & 0 \cr 0 & 1 \cr}$ and its restriction $\rho\vert$ to ${\cal L}_4$, such that $\nu\circ\rho=Id_{GL_4(\R)}$ and $\nu\vert\circ\rho\vert=Id_{{\cal L}_4}$. So $$GA_4(\R)=\R^4\odot GL_4(\R), \ {\cal P}_4=\R^4\odot{\cal L}_4 \eqno{(20)}$$ with composition law $$(\lambda^\prime,g^\prime)(\lambda,g)=(\lambda^\prime+g^\prime\lambda,g^\prime g).\eqno{(20a)}$$ 
The above s.e.s.'s pass to s.e.s.'s of the corresponding Lie algebras: $$\matrix{0 & \longrightarrow & \R^4 & \buildrel{\tilde{\mu}}\over\longrightarrow & ga_4(\R) & \matrix{\buildrel{\tilde{\nu}}\over\longrightarrow\cr\buildrel{\tilde{\rho}}\over\longleftarrow\cr} & gl_4(\R) & \longrightarrow & 0 \cr & & Id\uparrow & & \uparrow\iota & & \uparrow\iota & & & \cr 0 & \longrightarrow & \R^4 & \buildrel{\tilde{\mu}\vert}\over\longrightarrow & p_4 & \matrix{\buildrel{\tilde{\nu}\vert}\over\longrightarrow\cr\buildrel{\tilde{\rho}\vert}\over\longleftarrow\cr} & l_4 & \longrightarrow & 0 \cr} $$ with $gl_4(\R)=\R(4)$, $ga_4(\R)=\R^4\odot gl_4(\R)$ with Lie product $$(\lambda^\prime,R^\prime)(\lambda,R)=(R^\prime\lambda-R\lambda^\prime,[R^\prime,R]),\eqno{(21)}$$ where $[R^\prime,R]$ is the Lie product in $gl_4(\R)$ and $[\lambda^\prime,\lambda]=0$ in $\R^4$, $\tilde{\mu}(\xi)=(\xi,0)$, $\tilde{\nu}(\xi,R)=R$, and $\tilde{\rho}(R)=(0,R)$. $\tilde{\mu}$, $\tilde{\nu}$  and $\tilde{\rho}$ (and their corresponding restrictions $\tilde{\mu}\vert$, $\tilde{\nu}\vert$ and $\tilde{\rho}\vert$) are Lie algebra homomorphisms, with $\tilde{\nu}\circ\tilde{\rho}=Id_{gl_4(\R)}$ and $\tilde{\nu}\vert\circ\tilde{\rho}\vert=Id_{l_4}$. The s.e.s.'s split only at the level of vector spaces i.e. if $(\lambda,R)\in ga_4(\R)$, then $(\lambda,R)=\tilde{\mu}(\lambda)+\tilde{\rho}(R)$, but $(\lambda,R)\neq \tilde{\mu}(\lambda)\tilde{\rho}(R)$.

\

If ${\cal F}_{M^4}: GL_4\to FM^4\buildrel{\pi_F}\over\longrightarrow M^4$ and ${\cal A}_{M^4}: GA_4\to AM^4\buildrel{\pi_A}\over\longrightarrow M^4$ are respectively the bundles of linear and affine frames over $M^4$, where $FM^4= \cup_{x\in M^4}(\{x\}\times(FM^4)_x)$ with $(FM^4)_x$ the set of ordered basis $r_x= (v_{1x},\dots,v_{4x})$ of $T_xM^4$, and $AM^4=\cup_{x\in M^4}(\{x\}\times AM^4_x)$ with $AM^4_x=\{(v_x,r_x), \ v_x\in A_xM^4, \ r_x\in (FM^4)_x\},$ where $A_xM^4$ is the tangent space at $x$ considered as an affine space (Appendix 3), then one has the following bundle homomorphism: 
$$\matrix{AM^4\times GA_4 & \matrix{\buildrel{\beta\times\nu}\over\longrightarrow\cr\buildrel{\gamma\times\rho}\over\longleftarrow \cr} & FM^4\times GL_4\cr \psi_A\downarrow & & \downarrow\psi_F \cr AM^4 & \matrix{\buildrel{\beta}\over\longrightarrow\cr\buildrel{\gamma}\over\longleftarrow\cr} & FM^4 \cr \pi_A\downarrow & & \downarrow\pi_F \cr M^4 & \buildrel{Id}\over\longrightarrow & M^4\cr}$$ where $\beta(x,(v_x,r_x))=(x,r_x), \ \gamma(x,r_x)=(x,(0_x,r_x)), \ 0\in T_xM^n, \ \psi_F((x,r_x),g)=(x,r_xg),$ and $$\psi_A((x,(v_x,r_x)),(\xi,g))=(x,(v_x+r_x\xi,r_xg)).\eqno{(22)}$$

\

A {\it general affine connection} (g.a.c.) on $M^4$ is a connection in the bundle of affine frames ${\cal A}_{M^4}$; let $\omega_A$ be the 1-form of this connection, then $\omega_A\in \Gamma(T^*AM^4\otimes ga_4)$. 
From the smoothness of $\gamma$, the pull-back $\gamma^*(\omega_A)$ is a $ga_4$-valued 1-form on $FM^n$: $$\gamma^*(\omega_A)=\varphi\odot\omega_F,\eqno{(23)}$$ where $\omega_F$ is a connection on $FM^4$, and $\varphi$ is an $\R^4$-valued 1-form. There is a 1-1 correspondence between g.a.c.'s on $AM^4$ and pairs $(\omega_F,\varphi)$ on $FM^4$: $$\{\omega_A\}_{g.a.c.}\longleftrightarrow\{(\omega_F,\varphi)\}.\eqno{(24)}$$ 

\

$\omega_A$ is an {\it affine connection} (a.c.) on $M^4$ if $\varphi$ is the soldering (canonical) form $\theta_{FM^4}$ (see Appendix 2) on $FM^4$. Then, if $\omega_A$ is an a.c. on $AM^4$, $$\gamma^*(\omega_A)=\theta_{FM^4}\odot\omega_F. \eqno{(25)}$$ There is then a 1-1 correspondence $$\{\omega_A\}_{a.c.}\longleftrightarrow\{\omega_F\},\eqno{(26)}$$ since $\theta_{FM^4}$ is fixed. Also, if $\Omega_A$ is the curvature of $\omega_A$, then $$\gamma^*(\Omega_A)=D^{\omega_F}\theta\odot\Omega_F=T_F\odot\Omega_F \eqno{(27)}$$ since $D^{\omega_F}\theta_{FM^4} =T_F$: the {\it torsion} of the connection $\omega_F$ on $FM^4$.

\

We now consider the following diagram of bundle homomorphisms:

\

$$\matrix{AM^4\times GA_4 & \buildrel{\iota\times\iota}\over\longleftarrow & A^PM^4\times{\cal P}_4 & \matrix{\buildrel{\beta\vert\times\nu\vert}\over\longrightarrow\cr\buildrel{\gamma\vert\times\rho\vert}\over\longleftarrow\cr} & F^LM^4\times{\cal L}_4 & \buildrel{\iota\times\iota}\over\longrightarrow & FM^4\times GL_4\cr
 \psi_A\downarrow & & \psi_A\vert\downarrow & & \downarrow\psi_F\vert & & \downarrow\psi_F \cr AM^4 & \buildrel{\iota}\over\longleftarrow & A^PM^4 & \matrix{\buildrel{\beta\vert}\over\longrightarrow\cr\buildrel{\gamma\vert}\over\longleftarrow\cr} & F^LM^4 & \buildrel{\iota}\over\longrightarrow & FM^4\cr \pi_A\downarrow & & \pi_A\vert\downarrow & & \downarrow\pi_F\vert & & \downarrow\pi_F \cr M^4 & \buildrel{Id}\over\longrightarrow & M^4 & \buildrel{Id}\over\longrightarrow & M^4 & \buildrel{Id}\over\longrightarrow & M^4\cr}$$ where    $\pi_A\vert=\pi_P$, $\pi_F\vert=\pi_L$, $\psi_A\vert=\psi_P$ and $\psi_F\vert=\psi_L$, where $\psi_P$ and $\psi_L$ are the group actions in the Poincar\'e and Lorentz bundles, respectively.

\

The facts that $A^PM^4$ is a subbundle of $AM^4$ and $F^LM^4$ is a subbundle of $FM^4$, with structure groups and Lie algebras the corresponding subgroups and sub-Lie algebras, and the existence of the restrictions $\beta|:A^PM^4\to F^LM^4$ and $\gamma|:F^LM^4\to A^PM^4$, allow us to obtain similar conclusions for the relations between affine connections on the Poincar\'e bundle and linear connections on the Lorentz bundle: 

\

There is a 1-1 correspondence between affine Poincar\'e connections $\omega_P$ on $F^PM^4$ and Lorentz connections on $F^LM^4$: $$\{\omega_P\}\longleftrightarrow \{\omega_L\} \eqno{(28)}$$ with $$\gamma\vert^*(\omega_P)=\theta_L\odot\omega_L \eqno{(29)}$$ where $\theta_L=\theta_{FM^4}\vert_{F^LM^4}$ is the canonical form on $F^LM^4$. Also, $$\gamma\vert^*(\Omega_P)=D^{\omega_L}\theta_L\odot\Omega_L=T_L\odot\Omega_L.\eqno{(30)}$$ So, there is a 1-1 correspondence between curvatures of affine connections on $F^PM^4$ and torsion and curvature pairs on $F^LM^4$: $$\{\Omega_P\}\longleftrightarrow \{(T_L,\Omega_L)\}.\eqno{(31)}$$ For pure gravity governed by the Einstein-Hilbert action, $T_L=0$. 

\

5.2. {\it Local analysis: invariance of the actions $S_G$ and $S_{D-E}$}

\

To explicitly prove the Poincar\'e gauge invariance of GR and E-C theory, we have to consider as gauge transformations both the Lorentz part, already studied in the previous section, and the translational part. This last has to be done using the bundle of Poincar\'e frames ${\cal F}^P_{M^4}$. 

\

A gauge transformation or vertical automorphism in an arbitrary principal $G$-bundle $\xi:G\to P \buildrel{\pi}\over\longrightarrow B$, is a diffeomorphism $\alpha:P\to P$ such that i) $\alpha(pg)=\alpha(p)g$ and ii) $\pi(\alpha(g))=\pi(p)$, for all $p\in P$ and $g\in G$. Therefore, from ii), $\alpha(p)=pk$ for some $k\in G$. Then there is a bijection $Aut_{vert}(P)\buildrel{\Phi}\over\longrightarrow\Gamma_{eq}(P,G)$, $\Phi(\alpha)=\gamma_\alpha$ with $\alpha(p)=p\gamma_\alpha(p)$ and $\gamma_\alpha(pg)=g^{-1}\gamma_\alpha(p)g$; for the inverse, $\gamma\mapsto\alpha_\gamma$ with $\alpha_\gamma(p)=p\gamma(p)$.  

\

The action of ${\cal P}_4$ on $A^PM^4$ is given by $$\psi_P:A^PM^4\times{\cal P}_4\to A^PM^4, \ \psi_P((x,(v_x,r_x)),(\xi,h))\equiv(x,(v_x,r_x))(\xi,h)=(x,(v_x+r_x\xi,r_xh))$$ $$=(x,(v^\prime_x,r^\prime_x)),\eqno{(32)}$$ where $r_x=(e_{ax}), \ a=1,2,3,4,$ is a Lorentz frame, $h\in{\cal L}_4$, and $\xi\in\R^4\cong\R^{1,3}$ is a Poincar\'e gauge translation. For a pure translation, $h=I_L$ i.e. ${h_a}^b={\delta_a}^b$ and therefore $$(x,(v_x,r_x))(\xi,I_L)=(x,(v_x+r_x\xi,r_xI_L))=(x,(v_x+r_x\xi,r_x))$$ i.e. $$r^\prime_x=r_x.\eqno{(33)}$$ Therefore $e^\prime_{ax}=e_{ax}, \ a=1,2,3,4,$ and then, from the definition of $\omega^a_{\mu b}$ in section 2, $$\omega^{\prime a}_{\mu b}=\omega^a_{\mu b}\eqno{(34)}$$ since $\Gamma^\mu_{\nu\rho}=(\Gamma_{LC})^\mu_{\nu\rho}+K^\mu_{\nu\rho}$ remains unchanged (in the case of pure gravity $K^\mu_{\nu\rho}=0$). So the coordinate Ricci scalar $R$ is also a gauge scalar, and therefore $S_G$ is invariant.

\

By the same reason invoked in the case of Lorentz gauge invariance, $S_{D-E}$ is also invariant under translations: in an arbitrary $G$-bundle $P$ with connection $\omega$, a section $s$ of an associated bundle and its covariant derivative $D^\omega s$ transform in the same way.

\

The Poincar\'e bundle extends the symmetry group of GR and E-C theory to the semidirect sum $$G_{GR/E-C}={\cal P}_4\odot{\cal D},\eqno{(35)}$$ with composition law $$((\xi^\prime,h^\prime),g^\prime)((\xi,h),g)=((\xi^\prime,h^\prime)(g^\prime(\xi,h)g^{\prime -1}),g^\prime g).\eqno{(35a)}$$ The left action of ${\cal D}$ on ${\cal P}_4$ is given by the commutative diagram $$\matrix{A^PM^4 & \buildrel (\xi,h)\over\longrightarrow & A^PM^4 \cr g\downarrow & & \downarrow g \cr A^PM^4 & \buildrel (\xi^\prime,h^\prime)\over\longrightarrow \ & A^PM^4 \cr}$$ with $$g:A^PM^4\to A^PM^4, \ (x,(v_x^\mu{{\partial}\over{\partial x^\mu}}\vert_x,({e_{ax}}^\nu{{\partial}\over{\partial x^\nu}}\vert_x)))\mapsto (x,(v^{\prime\mu}_x{{\partial}\over{\partial x^{\prime\mu}}}\vert_x,({e_{ax}}^{\prime\nu}{{\partial}\over{\partial x^{\prime\nu}}}))),\eqno{(36)}$$ where $v_x^{\prime\mu}={{\partial x^{\prime\mu}}\over{\partial x^\alpha}}\vert_xv_x^\alpha$ and ${e_{ax}}^{\prime\nu}={{\partial x^{\prime\nu}}\over{\partial x^\beta}}\vert_x{e_{ax}}^\beta.$

\

{\bf 6. Gravitational potentials and interactions}

\

It is usually said that the coframes $e^a={e_\mu}^adx^\mu$ are the translational gravitational potentials (Hehl, 1985; Hehl et al, 1976; Hammond, 2002). This is {\it not} strictly true since these fields are not gauge potentials, but tensors, both in their Lorentz ($a$) and world ($\mu$) indices: see section 2 and (Hayashi, 1977). The translational gauge potentials are the 1-form fields ${B_\mu}^a$ locally defined as follows (Hayashi and Nakano, 1967; Aldrovandi and Pereira, 2007): $${B_\mu}^a={e_\mu}^a-{{\partial v_x^a}\over{\partial x^\mu}} \ \ or \ \ B^a=e^a-dv_x^a, \eqno{(37)}$$ where $v_x=\sum_{a=0}^3v_x^ae_{ax}\in A_xM^4$ (section 5.1.); the $v_x^a$'s are here considered the coordinates of the tangent space at $x$. A straightforward calculation leads to the following transformation properties:

\

Internal Lorentz: $${B_\mu}^{\prime a}={h_a}^b{B_\mu}^b-\partial_\mu({h_b}^a)v_x^b \ \ or \ \ B^{\prime a}={h_b}^aB^b-(d{h_b}^a)v_x^b, \eqno{(38)}$$

\

General coordinate transformations: $${B_\mu}^{\prime a}={{\partial x^\nu}\over{\partial x^{\prime\mu}}}{B_\nu}^a, \eqno{(39)}$$

\

Internal translations: $${B_\mu}^{\prime a}={B_\mu}^a-\partial_\mu \xi^a  \ \ or \ \ B^{\prime a}=B^a-d\xi^a. \eqno{(40)}$$ Then, $B=B_\mu dx^\mu={B_\mu}^adx^\mu b_a$, where $b_a$, $a=0,1,2,3$, is the canonical basis of $\R^4$, is the connection 1-form corresponding to the translations. 

\

In terms of the ${B_\mu}^a$ fields and the spin connection, the Ricci scalar (2) is given by $$R=({{\partial v_x^a}\over{\partial x^\mu}}{{\partial v_x^b}\over{\partial x^\nu}}+{{\partial v_x^a}\over{\partial x^\mu}}{B_\nu}^b+{{\partial v_x^b}\over{\partial x^\nu}}{B_\mu}^a+{B_\mu}^a{B_\nu}^b)(\partial^\mu\omega^\nu_{ab}-\partial^\nu\omega^\mu_{ab}+\omega^\mu_{ac}\omega^{\nu c}_b-\omega^\nu_{ac}\omega^{\mu c}_b). \eqno{(41)}$$ If one intends to use this Lagrangian density as describing a $({B_\mu}^a,\omega^\nu_{bc})$ (or $({e_\mu}^a,\omega^\nu_{bc})$) interaction (Randono, 2010), then immediately faces the problem that the ${B_\mu}^a$ (or ${e_\mu}^a$) does not have a free part (in particular a kinematical part), since all its powers are multiplied by $\omega$'s or $\partial\omega$'s. So an interpretation in terms of fields interaction seems difficult, and may be, impossible. 

\

{\bf Acknowledgements}

\

This work was partially supported by the project PAPIIT-IN118609, DGAPA-UNAM, M\'exico. We thank Fernando Izaurieta for useful comments.

\

{\bf References}

\

1. Aldrovandi, R. and Pereira, J. G. (2007). {\it An Introduction to Teleparallel Gravity}, Instituto de F\'\i sica Te\'orica, UNESP, Sao Paulo, Brazil.

\

2. Carroll, S. (2004). {\it Spacetime and Geometry. An Introduction to General Relativity}, Addison Wesley, San Francisco. 

\

3. Feynman, R. P., Morinigo, F. B., and Wagner, W. G. (2003). {\it Feynman Lectures on Gravitation}, Westview Press, Boulder, Colorado.

\

4. Gronwald, F. (1997). Metric-affine gauge theory of gravity. I. Fundamental structure and field equations, {\it International Journal of Modern Physics} D{\bf 6}, 263-303.

\

5. Gronwald, F. (1998). A note on gauge covariant translations in the gauge approach to gravity, {\it Acta Physica Polonica B} {\bf 29}, 1121-1129.

\

6. Hammond, R. T. (2002). Torsion gravity, {\it Reports on Progress in Physics} {\bf 65}, 599-649.

\

7. Hayashi, K. (1977). The gauge theory of the translation group and underlying geometry, {\it Physics Letters} {\bf 69}B, 441-444.

\

8. Hayashi, K. and Nakano, T. (1967). Extended Translation Invariance and Associated Gauge Fields, {\it Progress of Theoretical Physics} {\bf 38}, 491-507.

\

9.  Hehl, F. W., von der Heyde, P., Kerlick, G. D., and Nester, J. M. (1976). General relativity with spin and torsion: Foundations and prospects, {\it Review of Modern Physics} {\bf 48}, 393-416.

\

10. Hehl, F. W. (1985). On the Kinematics of the Torsion of Space-Time, {\it Foundations of Physics} {\bf 15}, 451-471.

\

11. Kobayashi, S. and Nomizu, K. (1963). {\it Foundations of Differential Geometry}, Volume I, J. Wiley, New York.

\

12. McInnes, B. T. (1984). On the affine approach to Riemann-Cartan space-time geometry, {\it Classical and Quantum Gravity} {\bf 1}, 115-123.

\

13. O'Raifearteigh, L. (1997). {\it The Dawning of Gauge Theory}, Princeton University Press, Princeton.

\

14. Randono, A. (2010). Gauge Gravity: a forward looking introduction, arXiv: gr-qc/1010.5822.

\

15. Smrz, P. K. (1977). Translations as Gauge Transformations, {\it Journal of Australian Mathematical Society} {\bf 20}, 38-45.

\

{\bf Appendix 1}

\

The Ricci scalar is given by $$R=\eta^{bd}{e_a}^\mu{e_d}^\nu(\partial_\mu\omega_{\nu b}^a-\partial_\nu\omega_{\mu b}^a+\omega_{\mu c}^a\omega_{\nu b}^c-\omega_{\nu c}^a\omega_{\mu b}^c)\equiv \eta^{bd}{e_a}^\mu{e_d}^\nu((\gamma)-(\delta)+(\alpha)-(\beta)),$$ with $(\gamma)=\partial_\mu\omega_{\nu b}^a$, $(\delta)=\partial_\nu\omega_{\mu b}^a$, $(\alpha)=\omega_{\mu c}^a\omega_{\nu b}^c$, and $(\beta)=\omega_{\nu c}^a\omega_{\mu b}^c$. 

\

Under the transformation $$\omega_{\mu c}^a={h_c}^l\omega_{\mu l}^{\prime r}{h^{-1}_r}^a+(\partial{h_c}^l){h^{-1}_l}^a$$ we have: 

\

$(\alpha)=(a)+(b)+(c)+(d)$ with $$(a)={h_c}^l\omega_{\mu l}^{\prime r}{h^{-1}_r}^a{h_b}^g\omega_{\nu g}^{\prime s}{h^{-1}_s}^c, \ (b)={h_c}^l\omega_{\mu l}^{\prime r}{h^{-1}_r}^a(\partial_\nu{h_b}^g){h^{-1}_g}^c,$$ $$(c)={h_b}^g\omega_{\nu g}^{\prime s}{h^{-1}_s}^c(\partial_\mu{h_c}^l){h^{-1}_l}^a, \ (d)=(\partial_\mu{h_c}^l){h^{-1}_l}^a(\partial_\nu{h_b}^g){h^{-1}_g}^c;$$ 

\

$(\beta)=(e)+(f)+(g)+(h)$ with $$(e)={h_c}^g\omega_{\nu g}^{\prime s}{h^{-1}_s}^a{h_b}^l\omega_{\mu l}^{\prime r}{h^{-1}_r}^c, \ (f)={h_c}^g\omega_{\nu g}^{\prime s}{h^{-1}_s}^a(\partial_\mu{h_b}^l){h^{-1}_l}^c,$$ $$(g)={h_b}^l\omega_{\mu l}^{\prime r}{h^{-1}_r}^c(\partial_\nu{h_c}^g){h^{-1}_g}^a, \ (h)=(\partial_\nu{h_c}^l){h^{-1}_l}^a(\partial_\mu{h_b}^g){h^{-1}_g}^c;$$

\

$(\gamma)=[1]+[2]+[3]+[4]$ with $$[1]={h_b}^n{h^{-1}_t}^a(\partial_\mu\omega_{\nu n}^{\prime t}), \ [2]=\omega_{\nu n}^{\prime t}\partial_\mu({h_b}^n{h^{-1}_t}^a), \ [3]=(\partial_\mu\partial_\nu{h_b}^n){h^{-1}_n}^a, \ [4]=(\partial_\nu{h_b}^n)(\partial_\mu{h^{-1}_n}^a);$$ and $(\delta)=[5]+[6]+[7]+[8]$ with $$[5]={h_b}^l{h^{-1}_s}^a(\partial_\nu\omega_{\mu l}^{\prime s}), \ [6]=\omega_{\mu l}^{\prime s}\partial_\nu({h_b}^l{h^{-1}_s}^a), \ [7]=(\partial_\nu\partial_\mu{h_b}^l){h^{-1}_l}^a, \ [8]=(\partial_\mu{h_b}^l)(\partial_\nu{h^{-1}_l}^a).$$ Now, 

\

$[3]-[7]=(\partial_\mu\partial_\nu{h_b}^n){h^{-1}_n}^a-(\partial_\nu\partial_\mu{h_b}^l){h^{-1}_l}^a=0,$

\

$(b)+(c)=\omega_{\mu l}^{\prime r}{h^{-1}_r}^a\partial_\nu{h_b}^l-\omega_{\nu g}^{\prime s}{h_b}^g\partial_\mu{h^{-1}_s}^a,$

\

$(f)+(g)=\omega_{\nu g}^{\prime s}{h^{-1}_s}^a\partial_\mu{h_b}^g-\omega_{\mu l}^{\prime r}{h_b}^l\partial_\nu{h^{-1}_r}^a;$

\

so

\

$((b)+(c))-((f)+(g))=\omega_{\mu l}^{\prime r}\partial_\nu({h^{-1}_r}^a{h_b}^l)-\omega_{\nu g}^{\prime s}\partial_\mu({h^{-1}_s}^a{h_b}^g);$

\

also,

\

$[2]-[6]=\omega_{\nu g}^{\prime s}\partial_\mu({h_b}^g{h^{-1}_s}^a)-\omega_{\mu l}^{\prime r}\partial_\nu({h_b}^l{h^{-1}_r}^a);$

\

then 

\

$((b)+(c))-((f)+(g))+([2]-[6])=0.$

\

Also,

\

$[4]-[8]=(\partial_\nu{h_b}^l)(\partial_\mu{h^{-1}_l}^a)-(\partial_\mu{h_b}^l)(\partial_\nu{h^{-1}_l}^a)$

\

and

\

$(d)-(h)=(\partial_\nu{h^{-1}_l}^a)(\partial_\mu{h_b}^l)-(\partial_\mu{h^{-1}_l}^a)(\partial_\nu{h_b}^l);$

\

so

\

$([4]-[8])+((d)-(h))=0.$

\

Finally, 

\

$[1]-[5]+(a)-(e)={h_b}^l{h^{-1}_s}^a(\partial_\mu\omega_{\nu l}^{\prime s}-\partial_\nu\omega_{\mu l}^{\prime s}+\omega_{\mu r}^{\prime s}\omega_{\nu l}^{\prime r}-\omega_{\nu r}^{\prime s}\omega_{\mu l}^{\prime r})$.

\

Therefore,

\

$R=\eta^{bd}{e_a}^\mu{e_d}^\nu{h_b}^l{h^{-1}_s}^a(\partial_\mu\omega_{\nu l}^{\prime s}-\partial_\nu\omega_{\mu l}^{\prime s}+\omega_{\mu r}^{\prime s}\omega_{\nu l}^{\prime r}-\omega_{\nu r}^{\prime s}\omega_{\mu l}^{\prime r})=\eta^{lt}{e_s}^{\prime\mu}{e_t}^{\prime\nu}(\partial_\mu\omega_{\nu l}^{\prime s}-\partial_\nu\omega_{\mu l}^{\prime s}+\omega_{\mu r}^{\prime s}\omega_{\nu l}^{\prime r}-\omega_{\nu r}^{\prime s}\omega_{\mu l}^{\prime r})$

\

$=R^\prime$.

\

{\bf Appendix 2}

\

The {\it soldering} or {\it canonical} form on the frame bundle ${\cal F}_{M^n}$ of an $n$ dimensional differentiable manifold, is the $\R^n$-valued differential 1-form on $FM^n$ given by 
$$\theta:FM^n\to T^*FM^n\otimes\R^n, \ (x,r_x)\mapsto\theta((x,r_x))=((x,r_x),\theta_{(x,r_x)}),$$ with
$$\theta_{(x,r_x)}: T_{(x,r_x)}FM^n\to\R^n, \ v_{(x,r_x)}\mapsto \theta_{(x,r_x)}(v_{(x,r_x)})=\tilde{r}_x^{-1}\circ d\pi_F|_{(x,r_x)}(v_{(x,r_x)})$$ i.e. 
$$\theta_{(x,r_x)}=\tilde{r}_x^{-1}\circ d\pi_F|_{(x,r_x)},$$ where $\pi_F$ is the projection in the bundle ${\cal F}_{M^n}:GL_n(\R)\to FM^n\buildrel{\pi_F}\over\longrightarrow M^n$ and $\tilde{r}_x$ is the vector space isomorphism 
$$\tilde{r}_x:\R^n\to T_xM, \ (\lambda^1,\dots,\lambda^n)\mapsto\tilde{r}_x(\lambda^1,\dots,\lambda^n)=\sum_{i=1}^n\lambda^iv_{ix}$$ with inverse
$$\tilde{r}_x^{-1}(\sum_{i=1}^n\lambda^iv_{ix})=(\lambda^1,\dots,\lambda^n).$$ In local coordinates $(x^\rho,X^\mu_\nu)$ on ${\cal F}_U$, $$\theta^\mu=\sum_{\nu=1}^n(X^{-1})^\mu_\nu dx^\nu$$ with $(X^{-1})^\mu_\nu(x,r_x)=(X^\mu_\nu(x,r_x))^{-1}=(v^\mu_{\nu x})^{-1}$, where $r_x=(v_{1x},\dots,v_{nx})$ and $v_{\nu x}=\sum_{\mu=1}^n v^\mu_{\nu x}{{\partial}\over{\partial x^\mu}}\vert_x$. Then $\theta^a={e_\mu}^a\theta^\mu={e_\mu}^a{(X^{-1})^\mu}_\nu dx^\nu={(X^{-1})^a}_\nu dx^\nu={e_\nu}^a dx^\nu=e^a$; so, if $\omega_F$ is a connection on ${\cal F}_{M^n}$, then $D^{\omega_F}\theta^a=d\theta^a+{\omega_F^a}_be^b=T_F^a$ is the torsion of $\omega_F$. 

\

{\bf Appendix 3}

\

An {\it affine space} is a triple $(V,\varphi,A)$ where $V$ is a vector space, $A$ is a set, and $\varphi$ is a free and transitive left action of $V$ as an additive group on $A$: $$\varphi:V\times A\to A, \ (v,a)\mapsto v+a,$$ with $$0+a=a \ and \ (v_1+v_2)+a=v_1+(v_2+a), \ for \ all \ a\in A \ and \ all \ v_1,v_2\in V.$$ Then, given $a,a^\prime\in A$, there exists a unique $v\in V$ such that $a^\prime=v+a$. Also, if $v_0$ is fixed in $V$, $\varphi_{v_0}:A\to A$, $\varphi_{v_0}(a)=\varphi(v_0,a)$ is a bijection.

\

Example. $A=V$: The vector space itself is considered as the set on which $V$ acts. In particular, when $V=T_xM^n$ and $A=T_xM^n$, the tangent space is called {\it affine tangent space} and denoted by $A_xM^n$. The points $``a"$ of $A_xM^n$ are the tangent vectors at $x$.

\end

[1] Cartan E. Sur une g\'eneralisation de la notion de courbure de Riemann et les espaces $\grave{a}$ torsion. Comptes Rendus Acad. Sci. 1922; 174: 593-595.

\

[2] Hehl FW, von der Heyde P, Kerlick GD, Nester JM. General relativity with spin and torsion: Foundations and prospects. Rev. Mod. Phys. 1976; 48: 393-416.

\

[3] von der Heyde P. The Equivalence Principle in the $U_4$ Theory of Gravitation. Lett. Nuovo Cimento. 1975; 14: 250-252.

\

[4] Hartley D. Normal frames for non-Riemannian connections. Class. Quantum Grav. 1995; 12: L103-105.

\

[5] Iliev BZ. Normal frames and the validity of the equivalence principle: I. Cases in a neighbourhood and at a point. J. Phys. A: Math. Gen. 1996; 29: 6895-6901.

\

[6] Nieto JA, Saucedo J, Villanueva VM. Relativistic top deviation equation and gravitational waves. Phys. Lett. A. 2003; 312: 175-186.

\

[7] Garcia de Andrade LC. Nongeodesic motion of spinless particles in the teleparallel gravitational wave background. 2002; arXiv: gr-qc/0205120.

\

[8] Landau LD, Lifshitz EM. The Classical Theory of Fields. 4th ed. Amsterdam: Elsevier; 1975: 259.  

\

[9] Carroll S. Spacetime and Geometry. An introduction to General Relativity. San Francisco: Addison Wesley; 2004: 106-108.

\

[10] Hehl FW. How does one measure torsion of space-time? Phys. Lett. A. 1971; 36: 225-226.  

\

[11] Pauli W. Theory of Relativity. New York: Dover; 1981: 163.

\

[12] Kosteleck\'y VA, Russell N, Tasson JD. Constraints on Torsion from Bounds on Lorentz Violation. Phys. Rev. Lett. 2008; 100: 111102.

\

[13] Fabbri L. On the Principle of Equivalence. 2009; arXiv: gr-qc/0905.2541.

\
 
[14] Fabbri L. On a Completely Antisymmetric Cartan Torsion Tensor. Annales de la Fondation Louis de Broglie. 2007; 32: 215-228; arXiv: gr-qc/0608090.

\

\end